# Proposal to measure with high precision
# π→e branching ratio and pion mean life

Vladimir I. Selivanov

*NRC "Kurchatov Institute", Moscow123182, Russia*
Email: selivanov@akado.ru

Experimental measurement of the decay ratio $R^{\pi}_{e/\mu} = \Gamma(\pi\to e\nu+\pi\to e\nu\gamma)/\Gamma(\pi\to\mu\nu+\pi\to\mu\nu\gamma) \approx 10^{-4}$ allows searching for interactions beyond the Standard Model: test of the hypothesis about the universality of the charged current $g_e=g_\mu=g_\tau$, search for new Pseudo-scalar Interactions. Currently, two collaborations are working in this area: PiENU (TRIUMF) and PEN at the Paul Scherrer Institute (PSI). Now $\sigma(R^{\pi,exp}_{e/\mu})$ is $1.87 \cdot 10^{-3}$. PEN collaboration strives to achieve accuracy of $\sigma\left(R^{\pi,exp}_{e/\mu}\right) = 5 \cdot 10^{-4}$. The proposed setup arose from the idea of separating with a high probability π→e and π→μ the decays in a plastic scintillator, where pions stop, and not measure the energy of decay positrons at all. An analysis of the proposed installation is given. It is estimated that an accuracy of $\sigma\left(R^{\pi,exp}_{e/\mu}\right) = 0.83 \cdot 10^{-4}$ can be achieved. The pion lifetime measured only for π→μ decay. Now $\sigma^{exp}(\tau_{\pi\to\mu}) = 1.9 \cdot 10^{-4}$ (measured in 1995). For the first time, a project proposed to measure the lifetime of a pion during π→e decay with an accuracy of $\sigma^{exp}(\tau_{\pi\to e}) = 0.7 \cdot 10^{-4}$.

## 1. Introduction

The most accurate test of the hypothesis of the Standard Model about the universality of the charged current $g_e/g_\mu = 1$ comes from the decay ratio $R^{\pi}_{e/\mu}=\Gamma(\pi\to e\nu+\pi\to e\nu\gamma)/\Gamma(\pi\to\mu\nu+\pi\to\mu\nu\gamma) \approx 10^{-4}$ [1]. Now $g_e/g_\mu = 0.9996 \pm 0.0012$ according to the results of the PiENU collaboration (TRIUMF) [2]. Maximum systematic error of this last record work occur when registering energy of decay positrons. The work approved in 2006 and the results published in 2015. Authors processing additional statistics and hope to achieve an accuracy of $\sigma(R^{\pi,exp}_{e/\mu}) = 10^{-3}$ [3]. The second work on measuring the probability of π→e decay was started in 2006 (PEN Collaboration, [4]). The work is not over. The authors hope to achieve an accuracy of $\sigma(R^{\pi,exp}_{e/\mu}) = 5\cdot 10^{-4}$ [5] after processing all the collected statistics. In this experiment, the largest systematic error also occurs when recording the energy spectrum of decay positrons. The proposed setup arose from the idea of separating with a high probability π→e and π→μ the decays in a plastic scintillator, where pions stop, and not measure the energy of decay positrons at all. This became possible after the development and implementation of silicon photomultipliers (SiPM) and, especially, after the development of SiPM with a large number of amplifying microcells (for example, KETEK company – SiPM with dimensions 3x3 mm$^2$, having 38800 microcells [6]).

## 2. Experimental Setup

The short lifetime of the pion (≈ 26 ns) and the required length of the pion channel of the accelerator lead to the fact that the intensity of the pion beam sufficient for the experiment (> $5\times10^5$ $\pi^+$/s) is possible at the momentum $p_\pi > 52$ MeV/s. Such a pion leaves ≈ 8 MeV in the scintillator, gives rise to ≈ $8\cdot 10^4$ photoelectrons (phe) in the plastic scintillator. With a quantum efficiency of 0.2 SiPM, this results in about 16000 SiPM cells firing. If the number of SiPM microcells is less than this number, the SiPM saturates and outputs the same signal for pion (8 MeV) and π + μ (8 + 4.12 ≈ 12 MeV), therefore a SiPM with a small number of cells (< 30,000) is not suitable for the proposed experiment. Attempts to reduce the number of photoelectrons lead to a situation where it is impossible to separate the cases π→e and π→μ→e with the necessary reliability due to the insufficient number of photoelectrons.

Using the SRIM program [7], many variants of the detector geometry, various scintillators, pion momentum magnitudes and momentum spread tested. The quality of the experiment (measurement accuracy) strongly (and not obvious) depends on the choice of the option due to the large number of critical parameters. For example, for $p_\pi > 56$ MeV/c ($\sigma = 1.3\%$), the events $\pi \to e$ and $\pi \to \mu \to e$ cannot be distinguished. Undoubtedly, it will be required to select the optimal variant using GEANT4 with the statistics $\approx 10^{12}$ events for each variant. One of these options described below. The proposed setup for further research shown in Fig. 1.

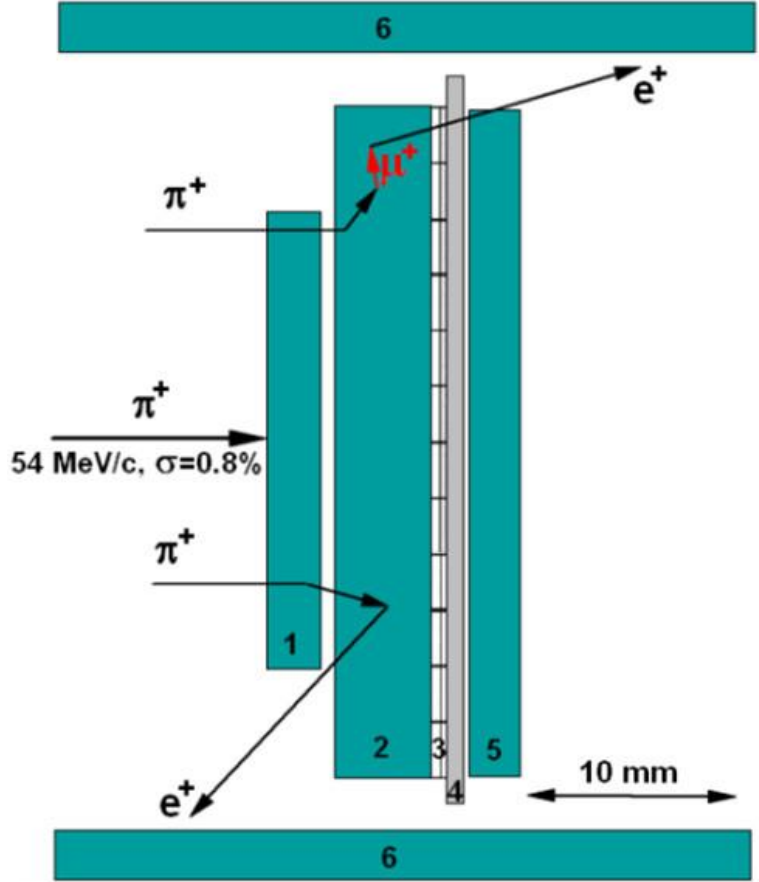

Fig. 1. Installation sketch. All scintillators 1, 2, 5, 6 are BC404 ($\tau_r = 0.7$ ns, $\tau_d = 1.8$ ns, 10500 photoelectrons/MeV); scintillator #1 - 3x24x24 mm$^3$. Scintillator #2: 5.5x36x36 mm$^3$. #3 - 12x12 = 144 SiPM PM3315-WB-CO [6]. #4 - PCB for signal routing from SiPM PM3315-WB-CO assembly. #5 - scintillator 3x36x36 mm$^3$. #6 - scintillator 3x40x40 mm$^3$ (4 pieces), they surround the installation from four sides. SiPM PM3325-WB-D0 [8] record signals from the ends of scintillators 1, 5, 6. SiPM parameters: PM3325-WB-D0 - overall dimensions 3.3x3.3x0.6 mm$^3$, sensitivity zone 3x3 mm$^2$, quantum efficiency 0.32, 13920 microcells; PM3315-WB-CO - overall dimensions 3.3x3.3x0.6 mm$^3$, sensitivity zone 3x3 mm$^2$, quantum efficiency 0.2, 38800 microcells. The installation has an angle of $\approx 2\pi$ for $\sigma(R_{e/\mu}^{\pi,\exp})$ measurement (without #1 and #6) or $\approx 3\pi$ (with #6) for $\sigma^{\exp}(\tau_{\pi \to e\nu})$ study. Fast outputs SiPM PM3315-WB-CO are combined and served by one ADC.

The distribution of pion stops in the target and their scattering shown in Figs. 2 and 3.

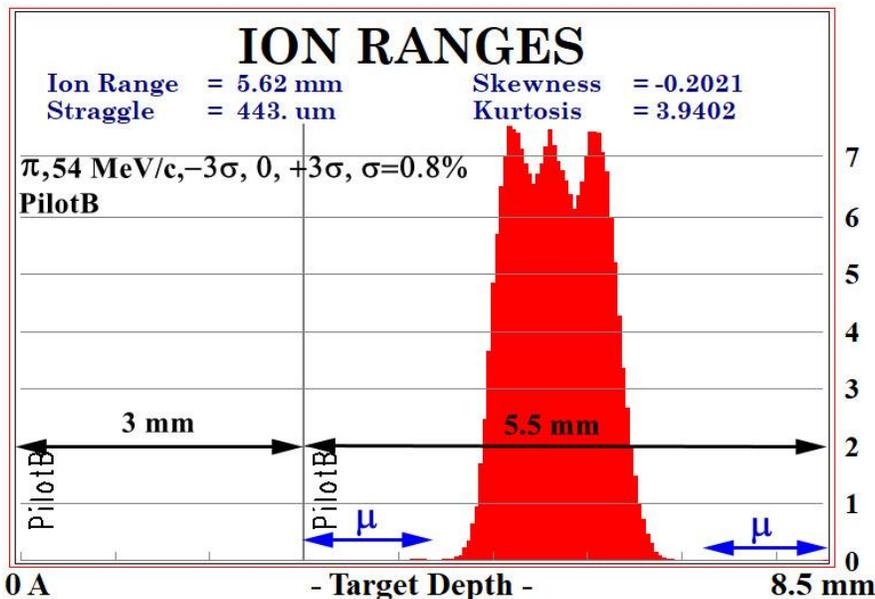

Fig. 2. Distribution of pion stops in the installation (Fig. 1). Three peaks correspond to pions with 54MeV/c - 3σ, 54MeV/c, and 54MeV/c + 3σ (σ = 0.8%). The blue arrows indicate the range of muons (≈ 1.3 mm) for π→μ decay.

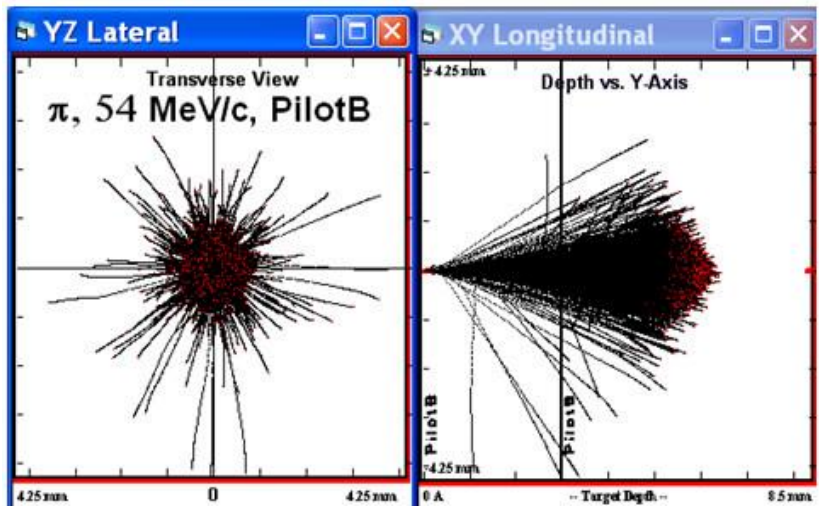

Fig. 3. Scattering of pions in the setup.

## 3. Analysis

### 3.1. Estimation technique

The setup analyzed as follows: using the SRIM program, pions and muons generated at momenta $p_0 - 3\sigma$, $p_0$, $p_0 + 3\sigma$, $p_0 = 54$ MeV/c and $\sigma = 0.8\%$. With such beam parameters, the required intensity of $\approx 5 \cdot 10^5 \pi^+$/sec is achieved at the PSI accelerator (Switzerland) [9]. Then the probability of error was calculated, i.e. the probability of false identification of the case in the worst conditions, for example, when identifying a pion and a muon from a beam in scintillator # 1, the minimum difference between the energies left by the particles was used. The same method applied for all cases. Obviously, when using GEANT4, the error probabilities will be lower, i.e. the maximum possible estimates given here. The estimates based on Gaussian distributions.

SiPM provides the ability to use either integral output (the signal is integrated within (0, 10) ns, 99% of photoelectrons are used) or fast output (formula 1). In this case, the maximum of signal is recorded. With the signal parameters real for BC404 – the leading edge $\tau_r = 1$ ns, the signal decay time $\tau_d = 2$ ns is used by 29% of the photoelectrons and the "error probability" value increases. In further evaluations, only the latter method was used.

The "error probability" due to the intrinsic amplitude resolution of the scintillator and the amplitude resolution of SiPM calculated separately. In the literature, various values of the intrinsic resolution of a plastic scintillator are given - from $\sigma(MeV) = 0.0072 \cdot \sqrt{E}$ (MeV) for pions 50 MeV [10] and worse. We use the expression $\sigma(MeV) = 0.031 \cdot \sqrt{E}$ (MeV) [11]. For SiPM we use the formula $\sigma = \sqrt{N_{fired}}$, where $N_{fired}$ is the number of fired microcells [12]. $N_{fired}$ nonlinearly depends on the number of incident photoelectrons (saturation SiPM).

Table 1. The probability of errors in identifying processes on the proposed installation.

| 54MeV/c, σ = 0.8%, FWHM = 1.9%, BC404 (3+5.5+3+4x3mm) → σE(MeV) = 0.031√E(MeV). SiPM → σ = √N$_{fired}$ 1. KETEK, 38800 microcells, 3x3 mm$^2$, PDE = 0.2 → #2, 8000phe/MeV; 2. SENSL, 14580 microcells, PDE = 0.32 → ##1, 5, and #6 (4 pieces), BC404, 5000phe/MeV. 29% phe of the total. | | | | | |
|---|---|---|---|---|---|
| | ## | | -3σ | 0 | +3σ |
| Beam momentum | 1 | 54 MeV/c | 52.7 | 54 | 55.3 |
| Initial energies of particle | 2 | π,MeV | 9.62 | 10.08 | 10.56 |
| | 3 | μ,MeV | 12.4 | 13.0 | 13.6 |
| Energy (MeV) left particles | 4 | π@#1(3mm) | 3.75 | 3.6 | 3.45 |
| | 5 | μ@#1(3mm) | 2.25 | 2.13 | 2.01 |
| | 6 | π@#2(5.5mm) | 5.87 | 6.48 | 7.11 |
| | 7 | μ@,#2(5.5mm) | 8.85 | 9.07 | 9.19 |
| | 8 | e@#2(5.5mm) | (0, 1.1) | | |
| Quantity of phe | 9 | e@##1,5,6(3mm) | (0, 990)→N$_{fired}$ = (0, 315) | | |
| Error probability | 10 | #1(π$_{min}$ vs μ$_{max}$), BC404 | 10$^{-9}$ | | |
| | 11 | #1(π$_{min}$ vs μ$_{max}$), SiPM | < 10$^{-10}$ | | |
| | 12 | #2, 5.5mm, (π$_{max}$ vs μ$_{min}$), BC404 | 3·10$^{-7}$ | | |
| | 13 | #2, 5.5mm, (π$_{max}$ vs μ$_{min}$), SiPM | 2·10$^{-10}$ | | |
| | 14 | #2, 5.5mm, BC404, π$_{min}$+μ(4.12MeV)+e(0, 1.1) MeV vs π$_{max}$+e(0, 1.1) MeV | < 4·10$^{-7}$, (t$_\pi$=0,t$_\mu$,t$_e$ < 2ns) | | |
| | 15 | #2, 5.5mm, SiPM, π$_{min}$+μ(4.12MeV)+e(0, 1.1MeV) vs π$_{max}$+e(0, 1.1) MeV | < 10$^{-9}$, (t$_\pi$=0, t$_\mu$, t$_e$ < 2 ns) | | |
| | 16 | #2, 5.5mm, BC404, μ(4.12MeV) vs e(0, 1.1) MeV | < 10$^{-15}$, (t$_e$-t$_\mu$ > 2 ns) | | |
| | 17 | #2, 5.5mm, SiPM, μ(4.12MeV) vs e(0, 1.1) MeV | < 10$^{-15}$, (t$_e$-t$_\mu$ > 2 ns) | | |

It is proposed to use 128 Channel Multihit TDC (100 ps/channel) V1190A-2eSST [13] on scintillators 1, 5, 6, and flash ADC [14] (or similar ADC) and TDC for scintillator #2.

## 3.2. Signal processing

Signal processing can be organized as follows. Fast outputs of several SiPM of each scintillator ##1, 2, 5, 6 are combined and fed to constant fraction discriminators (CFD) with a tracking threshold. At the fast output of the SiPM scintillator #1, two thresholds are set corresponding to the pion loss inside interval ≈ (3.45, 3.75) MeV (#4, Table 1), which reliably cuts off beam muons and positrons. The output signal serves as a "start" for all TDC. On discriminator from #5, 6 thresholds are set corresponding to an energy loss of (0.07, 1.3) MeV → (20, 370 phe). These signals serve as a "stop" for the corresponding TDCs. ADC distinguishes between muon and positron at $t_e - t_\pi \geq 2$ ns with an error probability $< 10^{-15}$, which is a negligible contribution of π→µ→e decay. Scintillator signal #1 opens TDCs in the interval (-10, +500) ns. Analysis of signals from ADCs allows separate construction of the spectra of π→µ→e and π→e decays. Separating the π→e statistics by π→µ→e statistics gives, minus the background at t < 0, the probability of π→e decay. Background value can be estimated by Monte Carlo with large statistics.

## 3.3. Systematics and statistics

From Table 1 (##10, 11, 12, 13) it can be seen that the probability of separation error (π vs. µ) in #1 is $10^{-9}$ and in #2 is $3 \cdot 10^{-7}$. This allows us not to estimate the error probability (π + e vs µ + e) in #2, since $\pi^+$ and $\mu^+$ are separated with a sufficient level of confidence.

According to #14, the appearance of all particles in the range ≤ 2 ns leads to the fact that the impurity of π→µ→e decays is $(4 \cdot 10^{-7}/1.2 \cdot 10^{-4}) = 3.3 \cdot 10^{-3}$ of the statistics of π→e decays. The Landau distribution for more energetic positrons π→e decays can increase the value. Such events are described as µ→e decays; therefore, all selected events of π→e decays should be described by the formula (1).

$$F_{\pi e}(t) - N_{bgrd} = N_0 \left( e^{-t/\tau_\pi} + a \cdot e^{-t/\tau_\mu} \right) \quad (1)$$

with parameters $N_0$, $\tau_\pi$, $\tau_\mu$, $a$, $N_{bgrd}$, where $N_{bgrd}$ = const., events at t < 0. If we fix $\tau_\pi$ in (1), then its error will give a systematic error $\sigma\left(R_{e/\mu}^{\pi,exp}\right) = 1.7 \cdot 10^{-4}$. Fixing $\tau_\mu$ gives an error of $5 \cdot 10^{-7}$, but the free parameter $\tau_\mu$ correlates with $\tau_\pi$ and gives an error ≈ $10^{-4}$. Therefore, in (1) only $\tau_\mu$ is fixed. The selected events of π→µ→e decays are described by a formula (2). ADC distinguishes between muon and positron at $t_e - t_\mu \geq 2$ ns with an error probability $< 10^{-15}$ (#16, Table 1), which is a negligible contribution of π→µ→e decay.

$$F_{\pi\mu e}(t) - N_{bgrd}^1 = N_0^1 \left( e^{-t/\tau_\mu} - e^{-t/\tau_\pi} \right). \quad (2)$$

$\tau_\mu$ is fixed in (2) also. The admixture of π→e decays in π→µ→e decays is obviously negligible. The ratio of two integrals $\int_{-10\,ns}^{500\,ns} A(t)\,dt / \int_{-10\,ns}^{500\,ns} B(t)\,dt$ gives the value of $R_{e/\mu}^{\pi,exp}$, where

$$A(t) = N_0 \cdot e^{-t/\tau_\pi}, B(t) = aN_0 \cdot e^{-t/\tau_\mu} + N_0^1 \left( e^{-t/\tau_\mu} - e^{-t/\tau_\pi} \right).$$

The decay positron identification is also unequivocally determined by the "stop" time TDC of scintillator #5 when measuring π→e branching ratio, and #5 or #6 when pion mean life study.

Let us consider the possibilities of separating events at different times of their arrival for real signals with a leading edge $\tau_r$ = 1 ns and a decay time $\tau_d$ = 2 ns. The pulse y (t) written in the form [15]:

$$y(t) = R\frac{\tau_r + \tau_d}{\tau_d^2} e^{-\frac{t}{\tau_d}}\left(1 - e^{-\frac{t}{\tau_r}}\right); \quad R\frac{\tau_r + \tau_d}{\tau_d^2} = 1 \ @ \ E = 7 \ MeV \qquad (1),$$

R is the total number of photoelectrons. The pulses y (t) were drawn for different particles at different times of their arrival. Below is the critical case at $t_\mu - t_\pi$ = 2ns when the ADC begins to distinguish between $\pi^+$ and $\mu^+$ signals.

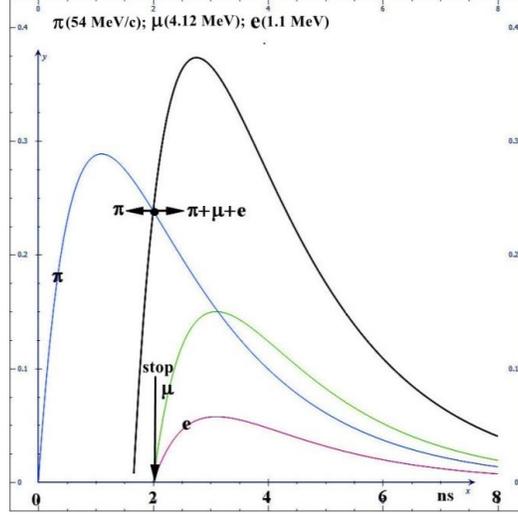

Fig. 4. Signals from particles in scintillator # 2. The μ-signal is the result of pion decay with $t_\mu - t_\pi$ = 2ns. At $t_\mu - t_\pi \leq$ 1.5 ns, the π and μ signals merge and the ADC perceives them as one signal. For $t_\mu - t_\pi \geq$ 2ns, the ADC registers both signals even if the muon decays instantly.

## 4. Discussion

The $\mu^+$ and $e^+$ impurities in the beam can significantly increase the probability of overlapping events. Perhaps reworking the pion channel, as done by PiENU [16], will contribute to a greater success of the proposed project.

Four #6 detectors are not used for decay spectrum acquisition. They only serve to reduce the cosmic background. This is due to the possibility of a positron escape with an energy loss of ≈7.2 MeV when crossing the entire scintillator #2. Such events worsen some of the "error probability" values of Table 1 up to a value of 1 and destroy the very idea of distinguishing between π→e and π→μ decays.

Obviously, the proposed method for assessing project capabilities is ideal. It does not take into account, for example, the scattering of particles in a scintillator (Fig. 3), the Landau distribution for decay positrons, etc. On the other hand, the maximum possible error probability is estimated. The data obtained allow us to hope to obtain quite acceptable estimates of the project with its further verification by the Monte Carlo method.

The π→eν positrons have energy of $E_e$ ≈ 70 MeV and, obviously, not absorbed in the setup (Fig. 1) until they recorded by detector #5. The spectrum of μ→e decay positrons can be represented as (for example, [17]) $N(\varepsilon) \sim (3 - 2\varepsilon)\cdot\varepsilon^2$, where $\varepsilon = E_e/E_{max}$, $E_{max}$ = 52.8 MeV. According to preliminary estimates, < $10^{-3}$ positrons of μ→e decay are absorbed in ##2, 3, 4, or have too low an energy for reliable registration in #5. This effect leads to a systematic shift in the probability value of π→eν. Its value will calculated with sufficient accuracy using GEANT4.

Cosmic background < $1s^{-1}$ falls on the installation. When the intensity of π→e decays is ≈ 5 · $10^5 s^{-1} \cdot 1.2 \cdot 10^{-4}$ = $60 s^{-1}$, the background value is ≈ 2%. For the most part, the background is cut of by coincidences #2 with any other detectors, since they cover an angle of ≈ 4π for external particles around scintillator #2.

Statistical errors for the selected observation interval of 500 ns. Intensity $\pi = 5 \cdot 10^5 s^{-1}$ (2μs between π), total = $5 \cdot 10^{12}$ for $10^7$sec → $6 \cdot 10^8$ π→e; 0.5/2 = 25% will be lost due to overlapping particles from the beam; 50% is lost due to the 2π registration angle; total – $2.25 \cdot 10^8$ π→e; statistical accuracy – $\sigma(R_{e/\mu}^{\pi,exp}) = 0.67 \cdot 10^{-4}$.

The lower figure shows the energy distributions in a plastic scintillator at a momentum of 72 Mev/c [18]. The drawing belongs to the authors of the publication [4].

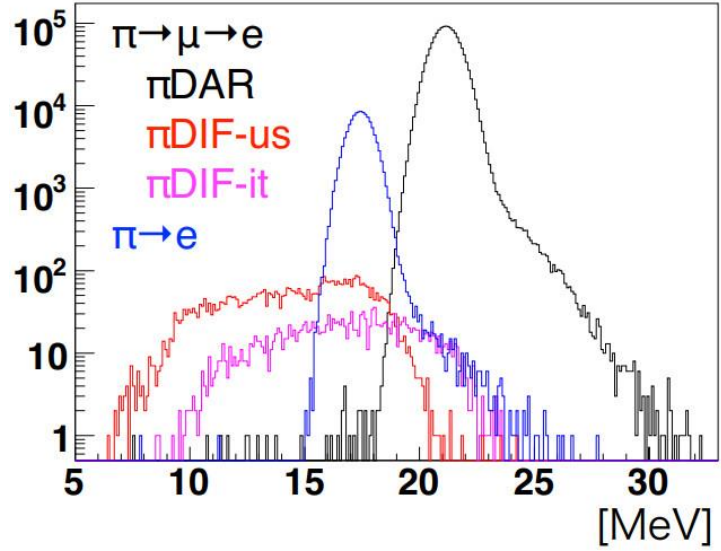

Fig. 5. Energy losses in the target during various processes. The "tails" of distributions towards higher energies arise due to the Landau distribution of energy losses by positrons. πDAR - pion decay-at-rest, πDIF-us - pion decay-in-flight before # 2, πDIF-it - pion decay-in-flight in # 2, μDIF - muon decay-in-flight in # 2.

A contribution to the systematics of $\pi^+$ and $\mu^+$ decays on the fly (NDIFπ→eν/Nπ→eν, NDIFπ→μν/Nπ→μν, and NDIFμ→eνν/Nμ→eνν) estimated in [19]. Their contribution is ($10^{-6}$, $10^{-5}$).

It should be noted that the same successful results are obtained at $p_\pi$ = 60 MeV/c, for example. Nevertheless, this requires $\sigma(p_\pi) \approx 0.3\%$. Appropriate assessments have been made. Estimates were made for the piE1 PSI beam [9] at $p_\pi$ = 54 MeV/c and $\sigma(p_\pi)$ = 0.8% (FWHM = 1.9%). Slightly worse results were obtained at $p_\pi$ = 52 MeV/c and $\sigma(p_\pi)$ = 1.3% (FWHM = 3.1%). According to (9), under these conditions, an intensity of $10^6$ $\pi^+$/sec is achieved.

## 5. Conclusion 1

Analysis shows that the probability of π→eν decay is determined with an accuracy of $\sigma(R_{e/\mu}^{\pi,exp})$ = $0.66 \cdot 10^{-4}$. We will assume that decays of particles on the fly and other reasons can introduce an additional systematic error of $\sigma\left(R_{e/\mu}^{\pi,syst}\right) = 0.5 \cdot 10^{-4}$. The contribution of the Landau distribution of the energy losses of positrons to the systematic errors is apparently small, but its value requires an accurate estimate using GEANT with large statistics. We conservatively expect to achieve an accuracy of $\sigma(R_{e/\mu}^{\pi,exp}) = 0.83 \cdot 10^{-4}$. This error is close to the theoretical uncertainty $\sigma(R_{e/\mu}^{\pi,theory}) = 10^{-4}$, calculated within the Standard Model [20]. Apparently, the value of the total error obtained by us can be reduced if we optimize by Monte Carlo the value of the recording interval, the beam intensity, detector geometry. In this project, we conservatively assume that $\sigma(R_{e/\mu}^{\pi,exp}) = 0.83 \cdot 10^{-4}$.

# 6. Measurement of the pion lifetime for π→e decay with an accuracy of 0.7·10⁻⁴

The pion lifetime measured only for π→μ decay. Now $\sigma^{exp}(\tau_{\pi\to\mu}) = 1.9\cdot 10^{-4}$ [21, 22] (measured in 1995). For the first time, a project proposed for measuring the lifetime of a pion during π→e decay with an accuracy of $\sigma^{exp}(\tau_{\pi\to e}) = 0.7\cdot 10^{-4}$. Generally speaking, $\sigma^{exp}(\tau_{\pi\to e}) = 10^{-4}$ can be obtained in the previous experiment. We will try to improve this accuracy, since it is not clear how to improve the accuracy $\sigma^{exp}(\tau_{\pi\to\mu})$ for 25 years.

The setting shown in Fig. 1 and all the above arguments remain valid.
Changes:
a) observation interval: 150 ns instead of 500 ns,
b) beam intensity: $10^6$ $\pi^+$/sec instead of $5\cdot 10^5$,
c) all four scintillators #6 are involved in this experiment.

Let us estimate the statistical errors, as is done in Table 2 under the selected conditions. We have $10^{13}$ pions in $10^7$ sec. Taking into account all losses, $6\cdot 10^{12}$ pions and $(6\cdot 10^{12})\cdot(1.2\cdot 10^{-4}) = 7.2\cdot 10^8$ π→e decays. Accuracy of $\sigma^{exp}(\tau_{\pi\to e\nu}) \approx 0.4\cdot 10^{-4}$. In this experiment, only selected cases of π→e decay are processed by formula (1). Choice of t = 0 does not change the value of $\tau_\pi$.

# 7. Conclusion 2

We will assume $\sigma^{exp}(\tau_{\pi\to e\nu}) = 0.7\cdot 10^{-4}$, since a complete analysis of systematic errors has not been done. It can be done with GEANT with a total statistics of $10^{13}$ events.